\documentstyle[eqsecnum,aps,epsfig]{revtex}
\begin{document}
\date{\today}
\title{
Chemical and Thermal Freeze-Out Parameters \\
from 1 to 200 A$\cdot$ GeV.}
\author{J. Cleymans\cite{present}}
\address
{Department  of  Physics,  University of Cape Town,
Rondebosch 7701, South Africa}
\author{K. Redlich$^{1,2}$}
\address{
$^1$Gesellschaft f\"ur Schwerionenforschung, D-64291 Darmstadt, Germany\\
$^2$Institute for Theoretical Physics, University of Wroc\l aw,
PL-50204  Wroc\l aw, Poland\\
}%
\maketitle
\begin{abstract}
The present knowledge about  hadrons produced in relativistic heavy
ion collisions is compatible with chemical freeze-out happening when the
energy density divided by the particle density reaches the value of 1
GeV.  This observation is used to determine
the energy dependence of the chemical freeze-out parameters $T_{ch}$ and
$\mu_B^{ch}$ 
for beam energies varying
between 1 and 200 A$\cdot$GeV. 
The consequences of this energy
dependence are studied for various particle ratios. Predictions for
particle ratios at beam energy 40 A$\cdot$GeV are presented.
The conditions for thermal freeze-out 
are also determined. 
These correspond either to an energy density of 45
MeV/fm$^3$ or to a particle density of 0.05/fm$^3$.
\end{abstract}
\section{Introduction}
It is the purpose of the present paper to 
determine the energy dependence of the chemical freeze-out parameters,
namely, the temperature $T_{ch}$ and the baryon chemical potential
$\mu_B^{ch}$ for beam energies in the 
range between 1 and 200
A$\cdot$ GeV \cite{heinz}. This covers the full range 
of energies available at the GSI/SIS, BNL/AGS and the
CERN/SPS experiments.  
These parameters determine the particle composition of the hadronic final
state. Below this temperature  inelastic collisions between hadrons
are no longer important and the hadronic abundances remain unchanged.
This information can then be used to make predictions
for beam energies that are in between those of the above experiments and
also, eventually, to extrapolate results to higher beam energies.

To determine in a relatively unambiguous 
way the chemical
freeze-out values in relativistic heavy ion collisions
it is best to use ratios of integrated particle yields. 
Such ratios are not very sensitive to the 
dynamics of the
underlying processes, one of the reasons 
being that integrated particle
yields are Lorentz invariant, i.e. a boost in the transverse direction
affects the momentum distribution of particles but not their number.
The restriction to
integrated particle yields therefore minimizes the 
model dependence of the freeze-out  parameters.
 This of course not true for particle yields 
 restricted to narrow kinematic
regions since these will be more dependent on the dynamics of the
process. 
During the past few years such analyses have been made for
the CERN/SPS 
\cite{stachel2,stachel3,stachel4,letessier,spieles,marek,NA44,gorenstein3}, 
the BNL/AGS \cite{stachel1,elliott,panagiotou,tounsi} and
the GSI/SIS \cite{keranen,prc} data. 
A discussion and review of the various results  can be found 
in Ref. \cite{sollfrank}.

The momentum distribution of
particles  produced in relativistic heavy ion collisions 
does not  show the dependence predicted by 
a statistical distribution.
The   longitudinal   direction   bears  no  resemblance  to  an
exponential, and  in 
the    transverse direction 
substantial deviations from an exponential fit with a universal slope
corresponding to the temperature exist.

In  section  2  we 
 discuss  the  
general case of hydrodynamic flow starting from the Cooper-Frye 
formula \cite{cooper-frye}.
As a particular case 
we  consider 
boost invariant longitudinal flow  \cite{bj}
 accompanied by
flow  in  the  transverse  direction.  We next discuss the case
where   the   longitudinal   flow  is  reproduced  by   a
superposition  of  fireballs.  
In all these cases particle
ratios  can be calculated 
as if they were given by a
static  Boltzmann  distribution. 

In section 3 we discuss the exact conservation of strangeness in
statistical models. This  is necessary  when
the number of strange particles is very small as is the case in e.g.
the GSI/SIS energy range. 

%
%
In section 4 we use the phenomenological 
observation that the freeze-out parameters all
correspond to an average energy per hadron of 1 GeV, independent of the
manner in which the system was created, to extract 
the energy dependence of $T_{ch}$ and $\mu_B^{ch}$.
This dependence is then used to track the behavior of various
hadronic ratios as a function of energy and to present predictions for a
beam energy of 40 GeV. 
We also determine the 
conditions for 
thermal freeze-out in the energy range between 1 A$\cdot$GeV and 200
A$\cdot$GeV. 
The 
values obtained previously 
\cite{NA49,nix,nix2,WA98flow,NA44flow,schnedermann,alam,shuryak,kaempfer}
are compatible with  an energy density $\epsilon\approx 45$
MeV/fm$^3$ 
or  a particle density 
$n\approx$ 0.05 fm$^{-3}$.
It is at present not possible to distinguish
between these two possibilities. 

In section 5 we summarize our results.
\section{Ratios of Thermal Particle Abundances}
In the thermal model particle densities are determined by a statistical
distribution,  denoted $n_i^0$ where the index $i$ refers to the
type of hadron, e.g $i = \pi^+, K^+, ...$ etc.
In the Boltzmann approximation 
$n_i^0$ is given by 
\begin{eqnarray}
n_i^0 &=& g\int {d^3p\over (2\pi)^3} e^{-(E_i-\mu_i)/T} \nonumber \\
      &=& gm_i^2TK_2(m_i/T) e^{\mu_i/T}
\label{eq:boltzmann} 
\end{eqnarray}
This expression applies for the density of particles inside a 
fireball at rest having a temperature $T$
and particle $i$ has a 
chemical potential $\mu_i$. 
If  the  system  is  in  chemical equilibrium then the possible
values  of $\mu_i$ are determined by the overall baryon chemical
potentials for baryon number, $\mu_B$, strangeness, $\mu_S$ and 
charge $\mu_Q$.
The value of $\mu_B$ is fixed by giving the overall baryon number
density $n_B$, and that of $\mu_S$ by fixing the overall strangeness
to zero, the value of $\mu_Q$ is fixed by giving the neutron surplus.
If the number of initial state neutrons is equal to the number of protons
as is the case in e.g. $S-S$ then $\mu_Q=0$, for $Pb-Pb$ $\mu_Q$ is
small and negative.
This means that for given values of $T$ and
$\mu_B$ one tunes the remaining parameters, $\mu_S$, and $\mu_Q$
in such a way as to ensure strangeness neutrality and the correct isospin
composition of the system.

At chemical freeze-out the particle density of hadrons of a given type 
is determined by  summing  over all hadronic densities
multiplied by the appropriate branching ratio, e.g. for $\pi^+$ one has
\begin{equation}
n_{\pi^+} = \sum_i n_i Br(i\rightarrow \pi^+).
\label{eq:branch}
\end{equation}
We have included 
in Eq. \ref{eq:branch} the contributions 
from all
particles and resonances with masses up to 2 GeV \cite{pdg}.
Thermal freeze-out happens when the elastic collisions between
hadrons cease.
Since it is obvious that the Boltzmann distribution is not
a good description of the momentum distribution of particles in
relativistic heavy ion collisions we will now 
discuss modifications which
lead to a more realistic description of the particle spectra.
\subsection{Cooper - Frye Formula}
In a hydrodynamic description taking into account flow in the 
longitudinal
and transverse directions, the final state particles will leave the
hadronic gas at freeze-out time.
The   momentum   distribution  of  particles  is  given  by  the
Cooper-Frye formula~\cite{cooper-frye}
\begin{equation}
E{dN\over  d^3p}  =  {g\over  (2\pi)^3}\int_{\sigma}  f(x,p)p^\mu
d\sigma_\mu,
\label{eq:cooper-frye}
\end{equation}
where the integration has to be performed over the freeze-out surface
described by $\sigma_\mu$. Its direction is perpendicular to the
surface and its magnitude is determined by the size of the freeze-out
surface. 
For the temperatures under consideration it is safe to neglect quantum
statistics and we will therefore work with the Boltzmann distribution
from now on; the generalization to Fermi-Dirac or Bose-Einstein
statistics is straightforward. We thus have
\begin{equation}
f(x,p) = \exp [(-p.u+\mu)/T],
\end{equation}
where $T$, $\mu$ and $u^\mu$ are the (space-time dependent) temperature,
  chemical   potential  and
four-velocity, respectively \cite{rischke}. 
As an example, for a static fireball, the
freeze-out surface is given by
\begin{equation}
d\sigma^\mu = (d^3x,\vec{0}),
\end{equation}
so that we obtain the expected result
\begin{equation}
E{dN\over  d^3p}  =  {gV\over  (2\pi)^3}E \exp [(-E+\mu)/T].
\end{equation}
After integrating the differential distribution over all momenta one
obtains the total number of particles \cite{heinz,rischke2}
\begin{eqnarray}
N &=& {g\over (2\pi)^3}\int{d^3p\over E} \int_{\sigma} f(x,p) p^\mu
d\sigma_\mu\nonumber \\
 &=& {g\over (2\pi)^3}\int_{\sigma}d\sigma_\mu\int{d^3p\over E}  f(x,p) p^\mu
\end{eqnarray}
for a Boltzmann distribution this becomes
\begin{equation}
N = {g\over (2\pi)^3} \int_{\sigma} d\sigma_\mu u^\mu \left[ 4\pi Tm^2
K_2\left({m\over T}\right)\right]
\end{equation}

And as a consequence :
\begin{equation}
{N_i\over N_j} = {n_i^0\over n_j^0} 
\end{equation}
Thus effects of hydrodynamic flow cancel out in the hadronic ratio,
provided there are unique freeze-out temperature and chemical potential.

\subsection{Rapidity plateau and  transverse expansion}
For 
boost-invariant cylindrical expansion along the
$z$-axis 
the correct variable to use is the proper time $\tau$
(recall $d^4x = \tau d\tau dy rdrd\phi $), we have
\begin{equation}
d\sigma^\mu = (\tau dy rdrd\phi,\hat{r}~\tau d\tau dy rd\phi),
\end{equation}
where the second component is in the $\hat{r}$ direction, i.e.
perpendicular to the surface of the cylinder. For the case where the
flow is azimuthally symmetric, i.e.  when an  average  is  made  over
all  events  or when only head-on collisions are considered, one has
therefore~\cite{ruuskanen,blaizot}
\begin{eqnarray}
\left( {dN_i\over dy m_tdm_t}\right)_{y=0}
&=& {g\over \pi} \int_\sigma r~dr~\tau_F(r)    \nonumber\\
& & \left\{ m_tI_0 \left( {p_t\sinh y_t\over T_{ch}} \right)\right.
          K_1 \left( {m_t\cosh y_t\over T_{ch}} \right) \nonumber \\
& & -\left( {\partial\tau_F\over\partial r} \right) p_t
          I_1 \left( {p_t\sinh y_t\over T_{ch}} \right)
    \left. K_0 \left( {p_t\cosh y_t\over T_{ch}} \right)
    \right\}e^{\mu_i/T_{ch}}
\end{eqnarray}
where $\tau_F(r)$ refers to the freeze-out time which in general depends
on $r$, so that the center of the cylinder freezes out before the
surface.

After integration over the transverse mass $m_T$ this leads to 
\begin{eqnarray}
\left( {dN_i\over dy }\right)_{y=0}
= & &{g\over \pi} \int_\sigma r~dr~\tau_F(r)    \nonumber \\
& & \left\{ \cosh(y_T) - \left({\partial\tau_F\over\partial r} \right) 
\sinh (y_T) \right\} m_i^2 T_{ch} K_2 \left( {m_i\over T_{ch}} \right)
e^{\mu_i/T_{ch}} .
\end{eqnarray}
If  the  freeze-out  temperature  is everywhere the same on the
freeze-out surface 
this  leads  to  the surprisingly simple result that the ratios
are unaffected by the hydrodynamic flow
\begin{eqnarray}
{\left( {dN_i/ dy }\right)_{y=0}\over
\left( {dN_j/ dy }\right)_{y=0}}
&= &{ m_i^2 T_{ch} K_2 \left( {m_i\over T_{ch}} \right)e^{\mu_i/T_{ch}}
\over m_j^2 T_{ch} K_2 \left( {m_j\over T_{ch}} \right)e^{\mu_j/T_{ch}}}
\nonumber\\
&=& {n^0_i\over n^0_j}
\end{eqnarray}
This is a case where the momentum distribution in no way resembles a
Boltzmann distribution : the rapidity distribution is flat and the
transverse momentum distribution is affected by transverse flow, yet 
the integrated ratios are the same as those of 
a static Boltzmann distribution.
\subsection{Superposition of Fireballs.}
In order to reproduce more closely the observed rapidity distribution 
it is useful to 
 consider  a superposition of fireballs 
along the 
rapidity axis \cite{sollfrank-heinz}. 
The resulting particle density (integrated
over
the particle rapidity $y$) is given by 
\begin{equation}
n_i  =
 \int_{-\infty}^{\infty}dy
          \int_{-Y}^YdY_{FB}~\rho(Y_{FB})
	   {dn_i^0\over dy}(y-Y_{FB})
\end{equation}
where $Y_{FB}$ is the position of the fireball and $\rho$ is the
distribution of fireballs and $Y$ is the largest value of the position in
rapidity space of a fireball. In the above integral it is possible to
interchange the integration limits so that one obtains
\begin{equation}
n_i  =
 n_i^0        \int_{-Y}^YdY_{FB}~\rho(Y_{FB})
\end{equation}
In a ratio the integral over the distribution of fireballs cancels out so
that one is left with
\begin{equation}
\displaystyle {n_i\over  n_j}   =  {n_i^0\over  n_j^0}
\end{equation}
With the conclusion that the ratio is the same as if it
were given by a  Boltzmann distribution. Note that the cancellation
is only possible if all fireballs are of a similar nature, i.e. they
should all have the same temperature.
%
%
%
\section{Exact Strangeness Conservation}
In relativistic heavy ion collisions  
the energy and  the number of hadrons in
the final state is
large enough to justify the use of the grand canonical ensemble.
 However if the temperature is very low, as is the case for the data from
 GSI/SIS, then the number of strange particles in the final state is very
 small and it becomes necessary to take this into account by using 
 exact strangeness 
 conservation
 \cite{redlich-hagedorn,old,Raf80,turko,mueller,derreth,zinovjev,crs}.
 The grand canonical ensemble only
 enforces strangeness conservation on average and 
 allows for fluctuations
 around strangeness zero. This is not good enough if the number of
 strange particles is very small. The same holds for applications of
 statistical descriptions to a small system like the ones created in 
 $p-p$ or $e^+-e^-$ collisions as has been done 
 recently in Ref. \cite{becattini}.
 
We therefore treat strangeness exactly, especially in the GSI/SIS energy
range but continue to use a grand canonical description 
for baryon number
and for charge since the number of baryons and charged particles is
always large.

To  restrict  the  ensemble  summation
to  a  fixed  value  of  the  strangeness $S$, 
one 
performs the following projection
\begin{equation}
Z_{S}={1\over 2\pi}\int_0^{2\pi}d\phi\ e^{-iS\phi}
\;  
Z(T,\lambda_B,\lambda_S,\lambda_Q)
\end{equation}
where the fugacity factor $\lambda_S$ has been replaced by :
\begin{equation}
\lambda_S = e^{i\phi}.
\end{equation}
and $Z$ is the standard grand canonical partition function.
Negecting for the moment the contribution of multi-strange particles
like, $\Xi$'s and $\Omega$'s one can write the
 partition function  as
\begin{equation}
 Z_{S=0}=
{1\over 2\pi}\int_0^{2\pi}d\phi~
\exp                       \left\{                      N_{0}
+N_{1}e^{i\phi}+N_{-1}e^{-i\phi}\right\}
\end{equation}
here $N_{1}$ stands for the sum of all single particle partition
functions with strangeness plus one:
\begin{equation}
N_{-1} \equiv N_{\Lambda} +N_{\bar{K}}+...
\end{equation}
while $N_{-1}$ stands for the sum of all single particle partition
functions with strangeness minus one:
\begin{equation}
N_{1} \equiv N_{\bar\Lambda} +N_{K}+...
\end{equation}
As an illustration we quote the explicit form of $N_{\Lambda}$ assuming
Boltzmann statistics
\begin{equation}
N_{\Lambda} \equiv {2V\over (2\pi)^3} \int d^3p \exp((-E_{\Lambda}+\mu_B)/T),
\end{equation}
It is clear from the above equation that baryon number is being treated
grand canonically. 

To calculate the partition function more explicitly we expand each term in
a power series:

\begin{equation}
 Z_{S=0}=
Z_0{1\over 2\pi}\int_0^{2\pi}d\phi~
\sum_{m=0}^{\infty}\sum_{n=0}^{\infty}{1\over m!}{1\over n!}
N_1^mN_{-1}^n\exp(im\phi)\exp(-in\phi),
\end{equation}
where $Z_0$ is the standard partition function for all particles 
having zero strangeness. 
Performing the integration over $\phi$ we are left with
\begin{equation}
 Z_{S=0}=
Z_0\sum_{n=0}^{\infty} {1\over n!^2}\left(N_1N_{-1}\right)^n
\end{equation}
where one recognizes the series expansion of the modified Bessel function
\begin{equation}
 Z_{S=0}=
Z_0I_0(x_1)
\end{equation}
where $x_1 \equiv 2\sqrt{N_1N_{-1}}$.
The particle densities can be deduced in the standard way from the
partition function. As an example we quote the number of kaons
\begin{equation}
N_K = Z_K {N_1\over\sqrt{N_1N_{-1}}}{I_1(x_1)\over I_0(x_1)}
\end{equation}
The total energy density can be calculated from
\begin{equation}
\left<E\right> = 
\left.T^2{\partial\over\partial T}\ln Z_{S=0}\right|_{\lambda_i=1}
\end{equation}
similarly the particle number is given by
\begin{equation}
\left<N\right> = 
T\left.{\partial\over\partial\mu_i}\ln Z_{S=0}\right|_{\lambda_i=1}
\end{equation}
Of special interest is the 
small volume limit (or, more correctly, the small particle number limit).
In this case the partition function can be expanded in a power series
in which case the integration over $\phi$ gives zero for all case where
the oscillating factors do not match, i.e.
\begin{eqnarray}
Z_{S=0}\approx 1 +
{1\over 2\pi}\int_0^{2\pi}d\phi
&&\left[ gV\int  {d^3p\over (2\pi)^3}e^{-E/T+i\phi}\right]\nonumber\\
\times &&\left[ gV\int  {d^3p\over(2\pi)^3}e^{-E/T-i\phi}\right]\nonumber\\
&&+\cdots
\end{eqnarray}
Thus it is clear that one needs
at  least  two  terms before one gets a non-vanishing result. For small
particle numbers this leads to the well-known suppression of strangeness.
\section{Chemical and Thermal  Freeze-Out Parameters.}
\subsection{Chemical Freeze-Out}
The considerations presented in the previous two sections make it possible  
to extract in a reliable
way the values of the chemical 
freeze-out parameters $T_{ch}$ and $\mu_B^{ch}$ from
ratios of integrated particle yields. 
This procedure has been followed repeatedly over the past few years
and 
results obtained 
from CERN/SPS, BNL/AGS and GSI/SIS data are 
shown in Fig. 1 and  in Table 1,  we
also show the results obtained in Ref. \cite{becattini} from
$e^+-e^-$ and $p-p$ collisions.
A low energy beam creates a hadronic gas with a corresponding low value
of $T_{ch}$ but a  high value of the baryon chemical potential.
This reflects the fact that such a system is predominantly made up of
nucleons and contains relatively few mesons.
At the other end, the CERN/SPS beam  creates a hadronic system
which has a high $T_{ch}$ but a very low value of $\mu_B^{ch}$, this is
because the hadronic system created is very rich in mesons and the
proportion of baryons diminishes accordingly.
It has been noted in Ref. \cite{prl} that, despite the
wide variation in beam energies, all these points have in common that the
energy density divided by the total particle density
is 1 GeV, independently of whether the system was created in $Ni-Ni$ at
1 A$\cdot$GeV or in $S-S$ at 200 A$\cdot$GeV. 
The energy density has been calculated using Eq. (3.11) and the particle
density was calculated using Eq. (3.12).
This means that, independently of how the
system was created
 from the moment the  average energy per
hadron becomes less than 1 GeV,  inelastic collisions cease 
to be important and the chemical
composition of the final hadronic state is fixed. We consider this to be
extremely remarkable.

Since the beam energy is increased smoothly as one follows the universal
freeze-out curve, Fig. 1, it is possible to determine the
energy dependence of the chemical freeze-out parameters, $T_{ch}$ and
$\mu_B^{ch}$, this will be done explicitly in the next section. We first
discuss some of the implications of the freeze-out curve.
\subsection{Non-Relativistic Nature of Hadronic Gas}
The phenomenological observation that the chemical freeze-out 
always corresponds
to an average energy per hadron of 1 GeV per hadron 
is easy to understand in the GSI/SIS energy range. Since
the low temperature 
makes the hadronic gas a non-relativistic one, nucleons dominate and only
very few other particles are present,  one can therfore use the
approximation
\begin{equation}
{\left<E\right>\over\left<N\right>} \approx  m_N + {3\over 2}T \approx 1 \rm{GeV}
\end{equation}
which immediately reproduces the value of 1 GeV per hadron.

At higher energies the hadronic gas becomes 
more and more a mesonic in nature and pions and $\rho$
mesons dominate. It is of interest to compute the average mass of 
particles that make up the gas. 
This mass decreases very slowly, starting  from the nucleon mass
to approximately the $\rho$-meson mass.
It is shown explicitly in Fig. 2. 
All hadrons in
the gas are thus to a  good approximation non-relativistic, except
for pions which dominate for low values of the baryon
chemical potential.  
The
non-relativistic character is further enhanced by looking at the value of
 the average energy along the 
freeze-out curve, this is shown 
in Fig. 2. Thus the following relation holds 
\begin{equation}
{\left<E\right> \over\left<N\right>}\approx  \left<M\right> 
+ {3\over 2}T \approx 1 \rm{GeV}  .
\end{equation}
The particle composition of the hadronic gas changes smoothly as one
follows the universal freeze-out curve from low to high temperatures.
At low temperatures the hadronic gas is dominated by nucleons as can be
seen explicitly in Fig. 3.
For high temperatures 
the hadronic gas is meson dominated, pions and rho
mesons making up the largest fraction.
\subsection{Thermal Freeze-Out Parameters}
After chemical freeze-out, the particle composition inside the hadronic
gas is fixed but elastic collisions still keep the system together until
the final, thermal freeze-out.
At this  stage  
the momentum distribution of
particles no longer changes and is then final.
The transverse momentum spectra therefore determine the
thermal freeze-out parameters. 
This has been done by several groups in the past few years
\cite{NA49,nix,nix2,WA98flow,NA44flow,schnedermann,alam,shuryak,kaempfer}
and  results are shown in Fig. 4.
We have to remember that systematic errors 
on the particle spectra can be substantial, this
introduces an additional uncertainty on the values of the parameters.
The baryon chemical potential at this stage cannot be determined directly. 
We
assume that the transition from chemical freeze-out to thermal freeze-out
proceeds in such a way that the entropy to baryon number remains conserved, ie.
$S/B$ = constant. These are indicated by dashed-dotted lines 
in Fig. 4. We note that
all results from 1 A$\cdot$GeV to 200 A$\cdot$GeV 
correspond to a fixed
energy density of 45 MeV/fm$^3$, i.e. if the energy density inside the
hadronic gas drops below this value, then the system ceases to exist
independently of how it was formed. We note that it is not
possible at present to distinguish between a fixed energy density and a
fixed particle density. Both curves are shown in Fig. 4.
\section{Discussion}
\subsection{Energy Dependence of Chemical Freeze-Out Parameters}
An inspection of Fig. 1
suggests  several possibilities for extracting the 
dependence of the thermal parameters on the beam energy.
It would e.g. be possible to plot the values of $T_{ch}$ as a function of
the beam energy and simply interpolate between them.
We follow a less direct method which relies on the use of results
contained in  Fig. 5. The values shown  in
this figure
have been  taken from the work of Gazd\'zicki \cite{marek2,marek3} which
uses 
a compilation of  experimental results on the pion
multiplicity divided by the number of spectator nucleons,
$\left<\pi\right>/A_{part}$. 
As can be seen from Fig. 5, in the relevant energy region  this ratio 
increases approximately linearly with the beam energy 
or, more precisely with the variable 
\begin{equation}
\sqrt{s}-\sqrt{s_{thr}}
\end{equation}
where $s_{thr}$ is the threshold energy for pion production given by
\begin{equation}
s_{thr}\equiv 2m_N+m_\pi.
\end{equation}
This approximate linear dependence provides  a
convenient parameterization for extracting the energy dependence. 
A polynomial interpolation was also used in order to test the sensitivity
of the parameterization.
To extract the dependence on beam energy we combine  Fig. 1
and Fig. 5. This is shown for one particular value in Fig.
6 where the crossing point between the two curves determines the value
of the 
chemical freeze-out parameters.
The resulting energy dependence of the freeze-out temperature $T_{ch}$ is
shown in Fig. 7, as one can see, the value, at first increases 
rapidly with beam energy and then converges
towards a maximum value of approximately 160 MeV.
We note that this value  is very close to the expected phase transition
to a quark-gluon plasma as indicated by results from lattice gauge theory
which give an upper limit of about 170 MeV for the critical
temperature \cite{laermann}.
The energy dependence
of the baryon chemical potential is shown in Fig. 8. 
In this case the value at first decreases rapidly and then tends towards
zero as the beam energy is increased.
\subsection{Hadronic Ratios.}
The  energy  dependence  obtained 
for $T_{ch}$ and $\mu_B^{ch}$ in the previous sub-section 
can be used to track
various particle ratios as the beam energy increases. 
Of particular interest are the ratios of strange to non-strange
particles since 
it is widely believed that these 
show  the largest deviation from chemical
equilibrium. Several analyses have indicated the need for an additional
parameter, $\gamma_S$, which measures the deviation from chemical
equilibrium of strange hadrons \cite{sollfrank,gammas}.
The $K^+/\pi^+$ ratio is shown in Fig. 9, it increases smoothly from the 
low  to  the higher energies and  reaches a maximum value
of  about  0.2  and  stays  approximately constant 
beyond this.   We   would  like  to  refer  to  these  values  as  the
\underbar{thermal model values for the $K^+/\pi^+$ ratio}. 
Comparing  these results with the experimental ones we see that
the  thermal  values nicely track the observed ratio in the GSI
and  BNL  energy  range.  
 Additional values of the $K^+/\pi^+$ ratio measured at mid-rapidity have
 become available recently from
BNL/AGS 
in the range 2 - 10.7 A$\cdot$GeV \cite{ogilvie}. 
All reported values are in  good
agreement 
with the thermal model values.
However  the  CERN/SPS  value  measured in  $S-S$
collisions is  clearly below  the  predicted  one. 
It has been argued recently \cite{tsukuba,gorenstein}
that this deviation is due to the formation of a quark-gluon plasma in
the initial state of the relativistic heavy ion collision.

 Next we consider the $K^+/K^-$ ratio. 
 This ratio is extremely sensitive to the beam energy
 because
 it varies by two orders of magnitude over 
 the energy range under consideration and is 
therefore a test of the thermal model. It is however not 
sensitive to the presence of $\gamma_S$ since it cancels out in the
ratio.
The comparison is shown in
 Fig. 10 and the agreement is good everywhere \cite{wang}. 
\subsection{Predictions}
The results 
presented in the previous sections 
substantially increase the 
predictive power of thermal models.
In the past it was not possible to use these models to calculate hadronic
ratios in a new energy domain.
In Table 2 we list  predictions of the model for a beam
energy of 40 A$\cdot$GeV and compare them to experimental results
obtained at CERN using $Pb-Pb$ at 158 A$\cdot$GeV. The main deviation 
can be found in ratios involving anti-baryons. This is because the baryon
chemical potential is predicted to be substantially larger at 40 GeV than
it 158 GeV, this suppresses all anti-baryons by an order of magnitude.
\section{Summary}
We have shown that all results obtained in relativistic heavy ion 
collisions for
beam energies between 1 A$\cdot$GeV and 200 A$\cdot$GeV
can be summarized in a very simple way: when the average energy per
hadron drops below 1 GeV, chemical freeze-out happens, the particle
composition of the final state is fixed, and inelastic collisions 
cease to be important. 
This happens for $Ni-Ni$ collisions at 1 A$\cdot$GeV, for $Pb-Pb$
collisions at 158 A$\cdot$GeV and for $S-S$ collisions at 200
A$\cdot$GeV. It is thus independent of the beam energy,  of the beam
particle and by implication, of the size of the hadronic volume.
Particle ratios for a beam energy of 40 A$\cdot$GeV have been given in
Table 2. We consider these predictions as highly reliable.

The next stage of the evolution of the hadronic system is thermal
freeze-out. Here the elastic collisions cease and the momenta
of the final state particles are fixed.
We have indicated that all available results are consistent
with the fact that this happens when the 
energy density drops below 45
MeV/fm$^3$
or when the 
particle density is less then 0.05/fm$^3$. It is not possible yet to
distinguish between these two possibilities
with the presently available data. 
\section*{Acknowledgments}

We acknowledge stimulating discussions with P. Braun-Munzinger,
B.  Friman,  M. Ga\'zdzicki, U. Heinz, W. N\"orenberg, H. Oeschler, H.
Satz and J. Stachel.
\newpage
%

%
\newpage
\section{Figure Captions}
\noindent Fig. 1 {{\it 
Chemical freeze-out parameters $T_{ch}$ and $\mu_B^{ch}$ obtained 
at   LEP, CERN/SPS, BNL/AGS
and GSI/SIS. References to the points are given in table 1.
The full line corresponds to an energy density over total particle
density of 1 GeV.
}}\\
Fig. 2 {{\it   
Average mass along the freeze-out curve of Fig. 1 as a function of the 
chemical freeze-out temperature.
}}\\
Fig. 3 {{\it 
Change of the particle composition of the hadronic gas 
along the freeze-out curve of Fig. 1 as a function of
the chemical freeze-out temperature. Shown are the fractions of nucleons
and of pions.
}}\\
%
Fig. 4 {{\it 
Thermal freeze-out (dashed and dashed-dotted lines) and chemical freeze-out 
(solid line) curves.
The
dotted lines connecting the chemical and thermal freeze-out curves correspond
to a fixed S/B (entropy/baryon number) ratio.
Only references to the
thermal freeze-out points are indicated explicitly. See Fig. 1 for
references to the chemical freeze-out points.
}}\\
Fig. 5 {{\it 
Average  number of pions divided by the number of participating
nucleons  as  a  function  of  the beam energy. The full line
corresponds   to   a   linear   fit,  the  dashed  line  is  a
polynomial interpolation.
}}\\
%
Fig. 6 {\it{ 
Intersection between the chemical freeze-out curve and  the line
corresponding to a given ratio of pions to A$_{part}$. The intersection
determines the values of $T_{ch}$ and $\mu_B^{ch}$ for the beam energy
corresponding to the indicated $\pi/A_{part}$ ratio.
References to the chemical freeze-out points are given in Fig. 1.
}}\\
Fig. 7 {{\it 
Variation of the temperature at chemical freeze-out as a
function of the beam energy.
}}\\
Fig. 8 {{\it 
Variation of the baryon chemical potential at chemical freeze-out as a
function of the beam energy.
}}\\
Fig. 9{{\it 
Dependence of $K^+/\pi^+$ 
ratio on beam energy along the chemical freeze-out curve of Fig. 1. 
Experimental results are indicated. See discussion in text.
}}\\
Fig. 10 {{\it 
Dependence of $K^+/K^-$ 
ratio on beam energy along the freeze-out curve of Fig. 1. Experimental
results are indicated.
}}\\
\newpage
\begin{figure}
\begin{center}
\epsfig{file=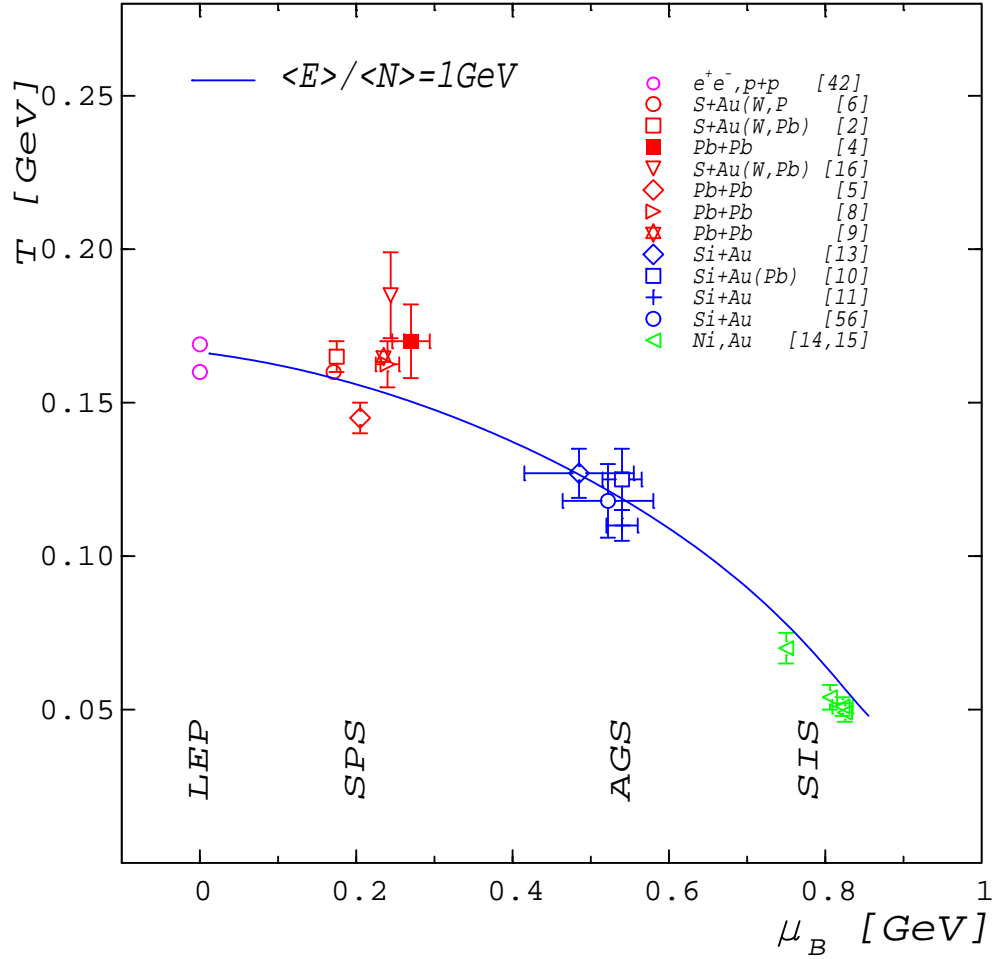}
\caption{{\it 
Chemical freeze-out parameters $T_{ch}$ and $\mu_B^{ch}$ obtained 
at   LEP, CERN/SPS, BNL/AGS
and GSI/SIS. References to the  points are given in table 1.
The full line corresponds to an energy density over total particle
density of 1 GeV.
}}
\label{fig1}
\end{center}
\end{figure}
\newpage
\begin{figure} 
\begin{center}
\epsfig{file=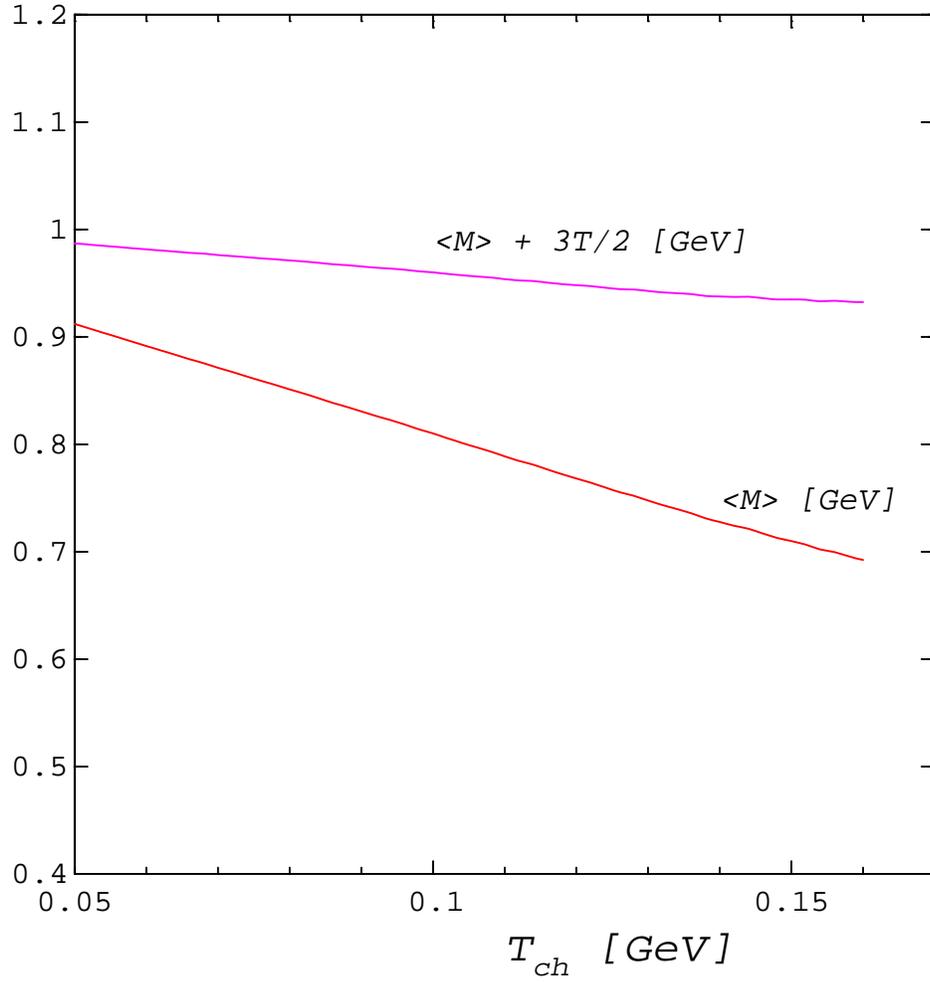}
\caption{{\it   
Average mass along the freeze-out curve of Fig. 1 as a function of the 
chemical freeze-out temperature.
}}
\label{fig2}
\end{center}
\end{figure}
\newpage
%
%
%
\begin{figure} 
\begin{center}
\epsfig{file=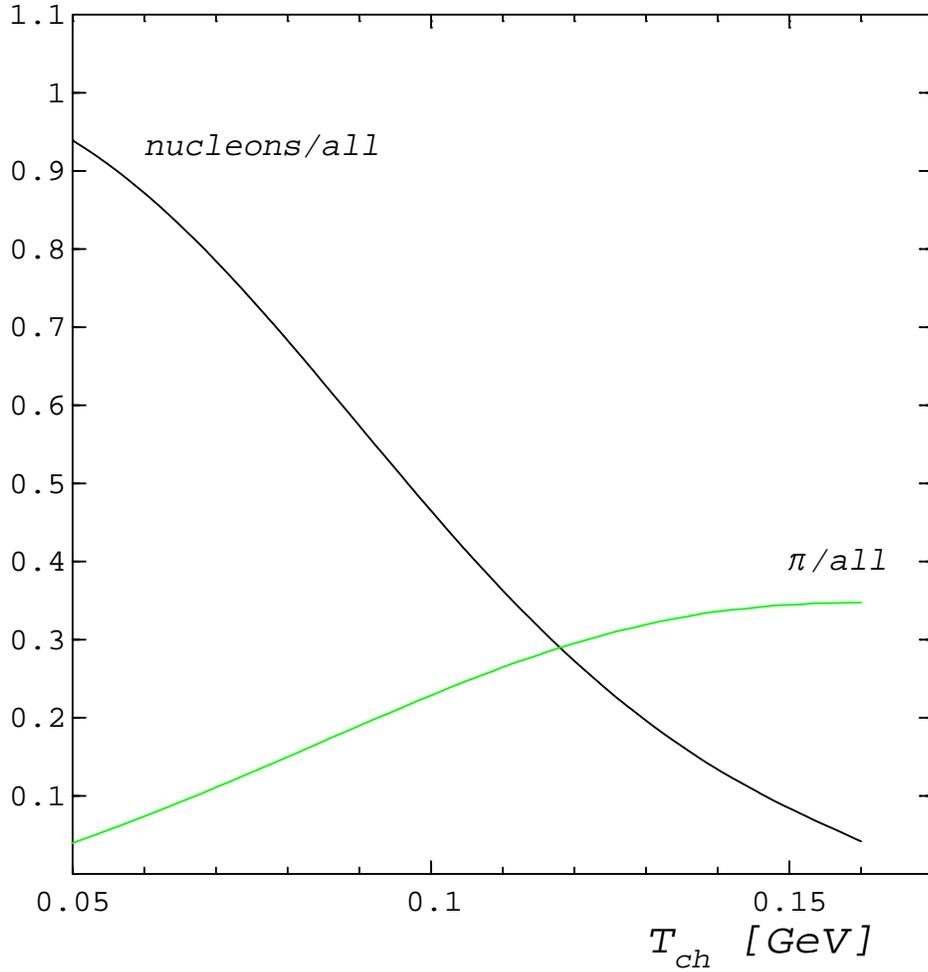}
\caption{{\it   
Change of the particle composition of the hadronic gas 
along the freeze-out curve of Fig. 1 as a function of
the chemical freeze-out temperature. Shown are the fractions of nucleons
and of pions.
}}
\label{fig3}
\end{center}
\end{figure}
\newpage
%

\begin{figure}
\begin{center}
\epsfig{file=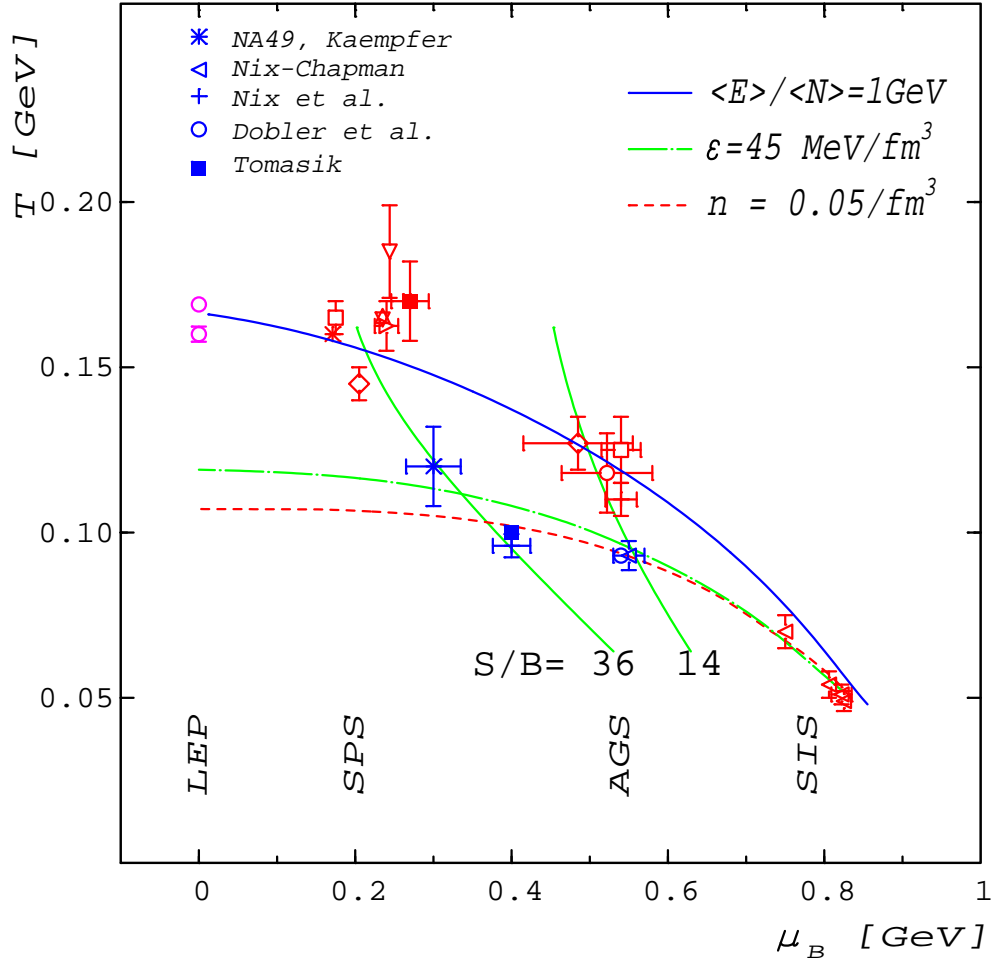}
\caption{{\it
Thermal freeze-out (dashed and dashed-dotted lines) and chemical freeze-out 
(solid line) curves.
The lines connecting the chemical and thermal freeze-out curves correspond
to a fixed S/B (entropy/baryon number) ratio. 
Only references to the
thermal freeze-out points are indicated explicitly. See Fig. 1 for
references to the chemical freeze-out points.
}}
\label{fig4}
\end{center}
\end{figure}
\newpage
\begin{figure}
\begin{center}
\epsfig{file=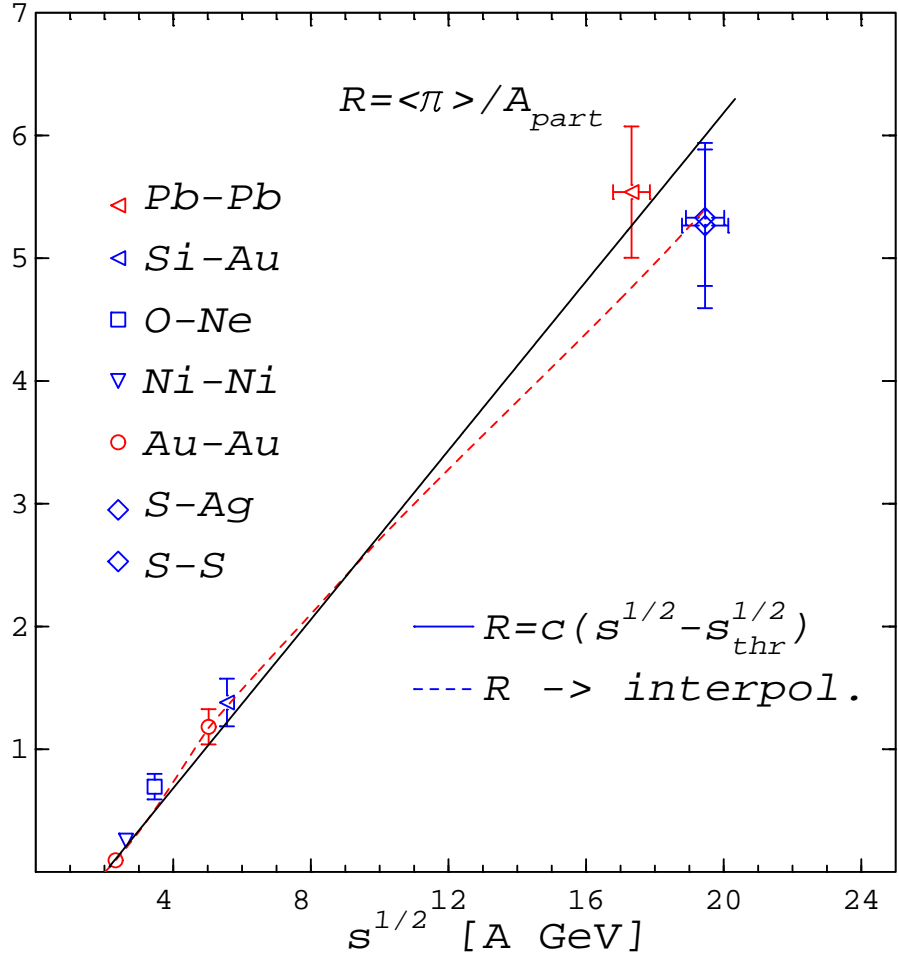}
\caption{{\it 
Average  number of pions divided by the number of participating
nucleons  as  a  function  of  the beam energy. The full line
corresponds   to   a   linear   fit,  the  dashed  line  is  a
polynomial interpolation.
}}
\label{fig5}
\end{center}
\end{figure}
\newpage
\begin{figure}
\begin{center}
\epsfig{file=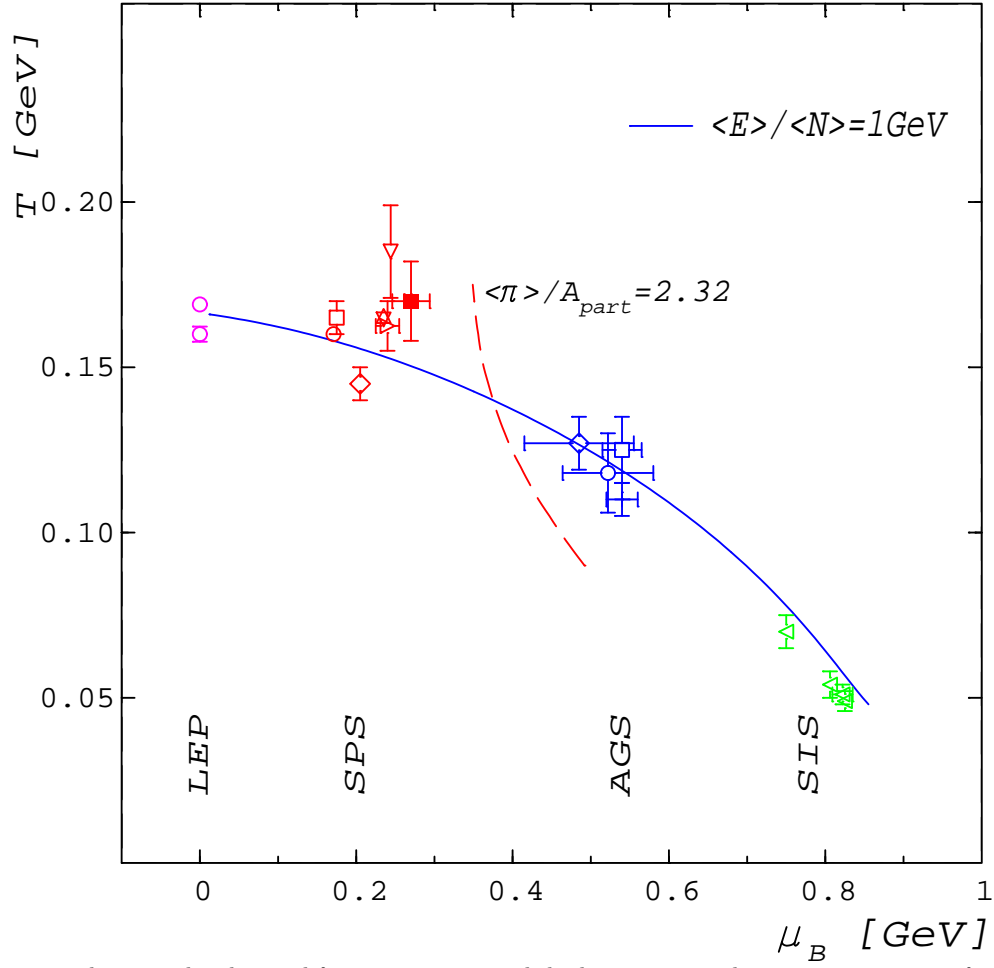}\\
\caption{{\it
Intersection between the chemical freeze-out curve and  the line
corresponding to a given ratio of pions to A$_{part}$. The intersection
determines the values of $T_{ch}$ and $\mu_B^{ch}$ for the beam energy
corresponding to the indicated $\pi/A_{part}$ ratio.
References to the chemical freeze-out points are given in Fig. 1.
}}
\label{fig6}
\end{center}
\end{figure}
\newpage
\begin{figure}
\begin{center}
\epsfig{file=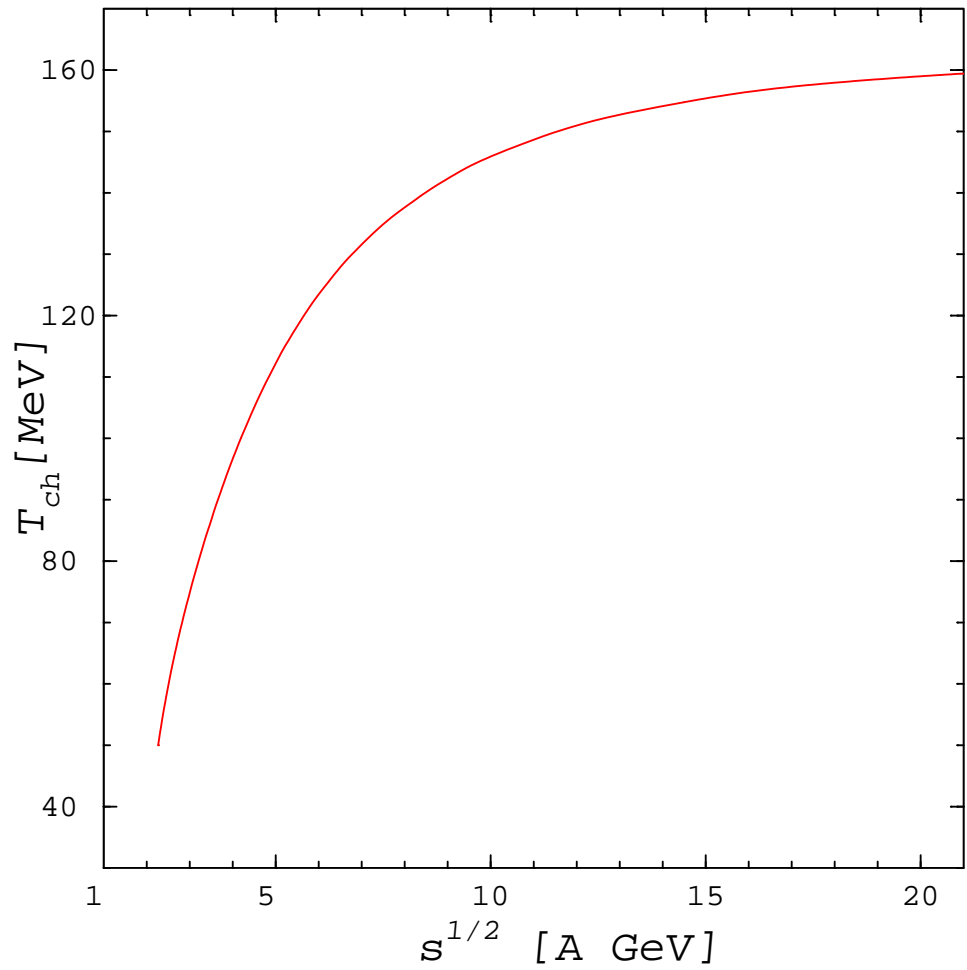}\\
\label{fig7}
\caption{{\it 
Variation of the temperature at chemical freeze-out as a
function of the beam energy.
}}
\end{center}
\end{figure}
\newpage
\begin{figure}
\begin{center}
\epsfig{file=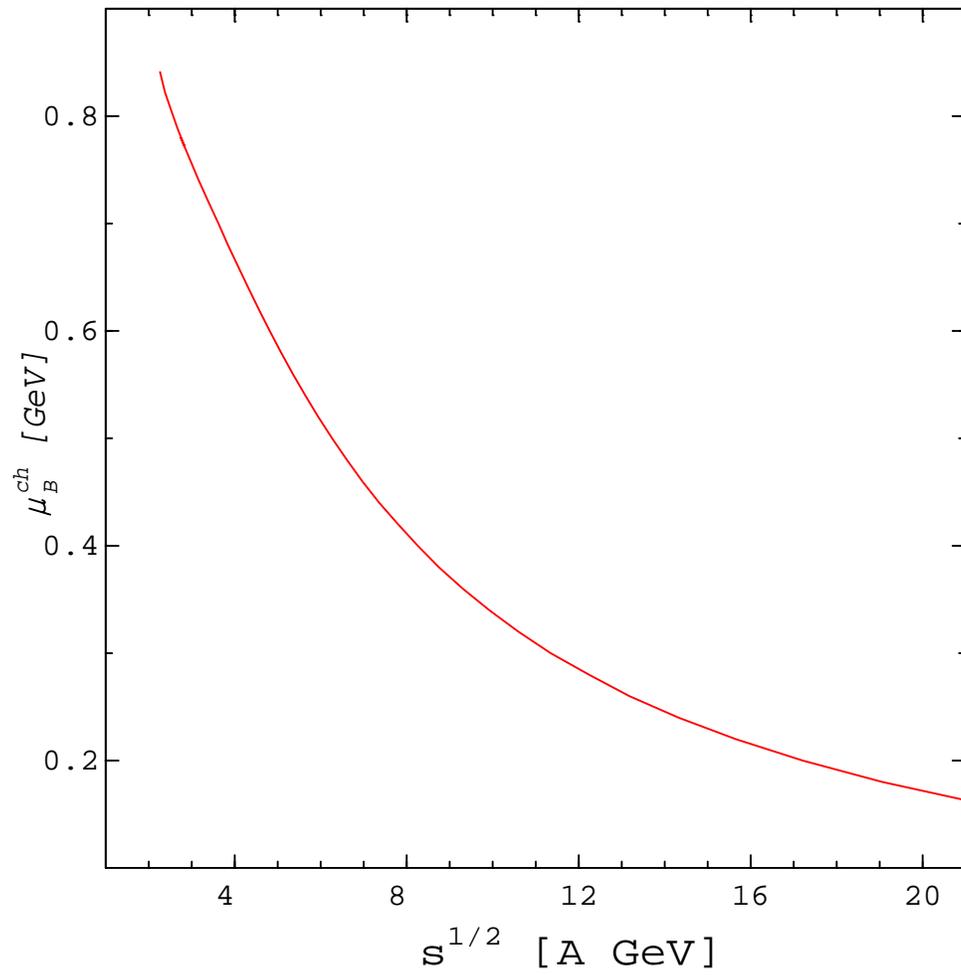}\\
\label{fig8}
\caption{\it{
Variation of the baryon chemical potential at chemical freeze-out as a
function of the beam energy.
}}
\end{center}
\end{figure}
\newpage
\begin{figure}
\begin{center}
\epsfig{file=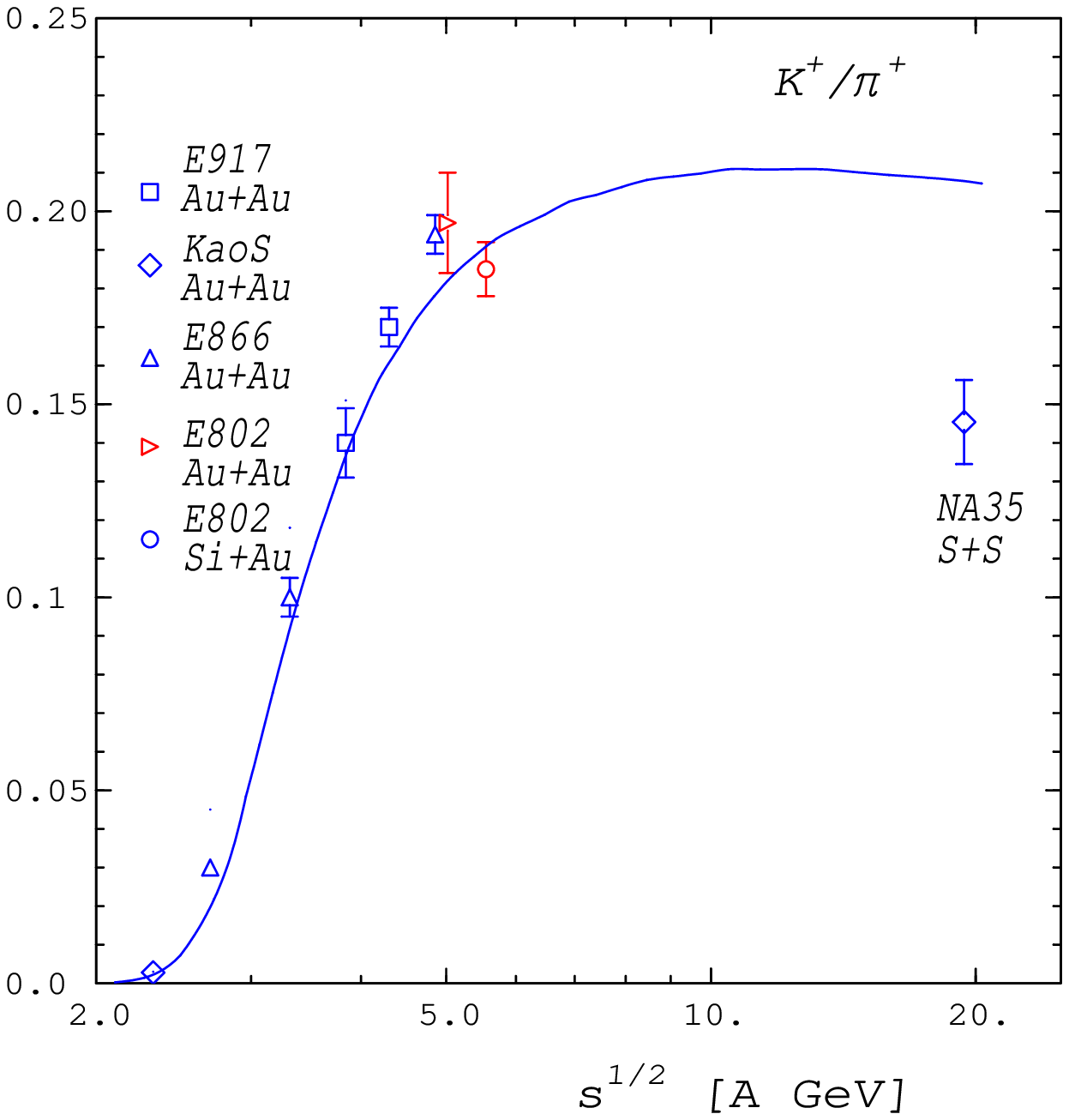}\\
\label{fig9}
\caption{\it{
Dependence of $K^+/\pi^+$ 
ratio on beam energy along the chemical freeze-out curve of Fig. 1. 
Experimental results are indicated. See discussion in text.
}}
\end{center}
\end{figure}
\newpage
%
\begin{figure}
\begin{center}
\epsfig{file=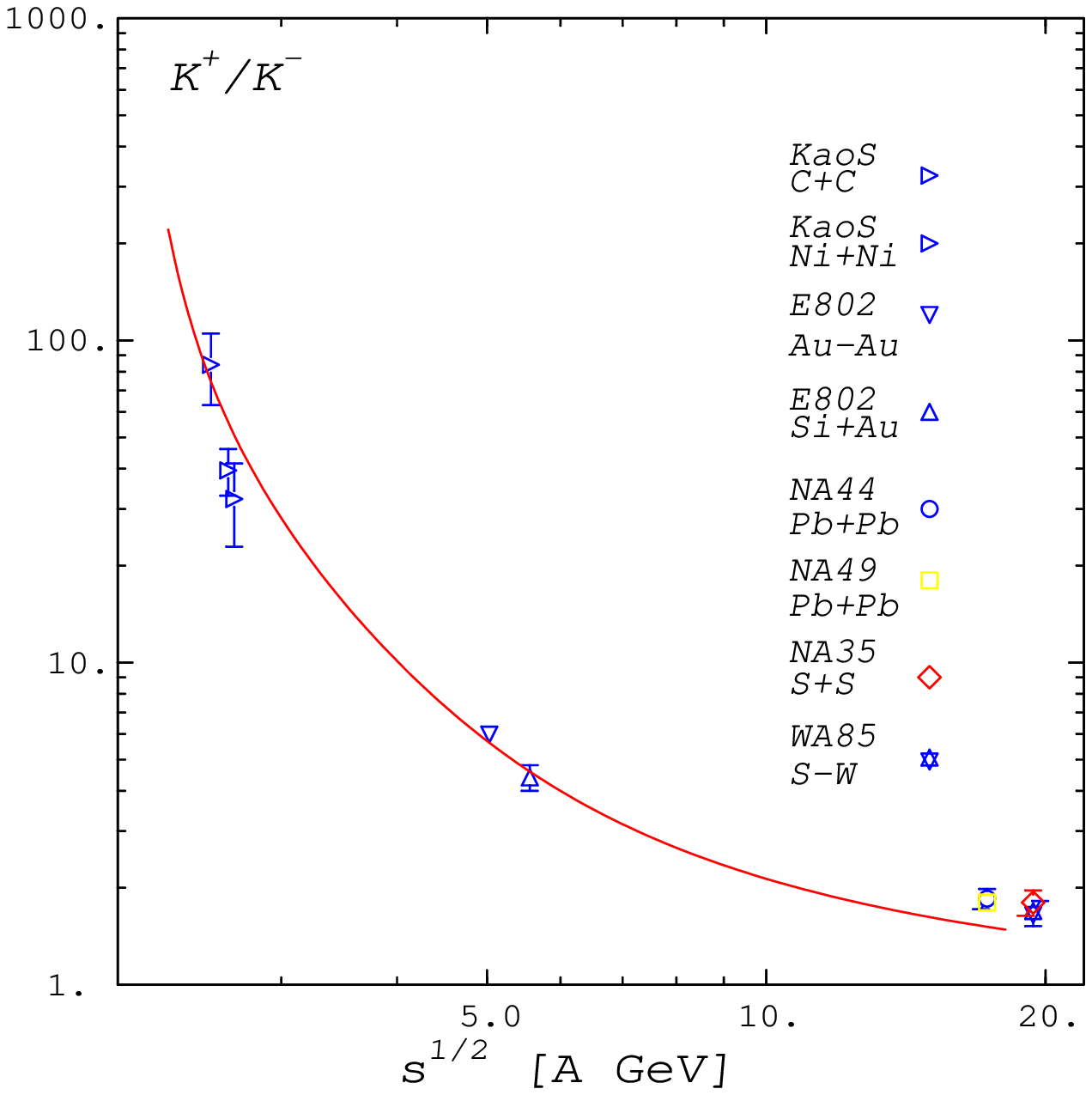}\\
\label{fig10}
\caption{\it{
Dependence of $K^+/K^-$ 
ratio on beam energy along the freeze-out curve of Fig. 1. Experimental
results are indicated.
}}
\end{center}
\end{figure}
\newpage
%
\begin{table}
\begin{center}
\begin{tabular}{|l|c|c|c|} 
Energy & $T_{ch}$ [MeV] &  $\mu_B^{ch}$ [MeV] & Reference\\
       &           &                &         \\ \hline \hline
SPS    &           &                &         \\ \hline
       &           &                &         \\
Pb + Pb 158 A$\cdot GeV$ &$170\pm 11$ &  $270\pm 24$& \cite{stachel4} \\ \hline
       &           &                &         \\
Pb + Pb 158 A$\cdot GeV$ &$165$ &  $235$& \cite{gorenstein3} \\ \hline
       &           &                &         \\
Pb + Pb 158 A$\cdot GeV$ &$145\pm 5$ &  $205\pm 5$& \cite{letessier} \\ \hline
       &           &                &         \\
S + S 200 A$\cdot$GeV&$180.5\pm 10.9$ &  $220.2\pm  18.0$& \cite{marek} \\ \hline
&                &&       \\
S + Ag 200 A$\cdot$GeV&$178.9\pm 8.1$ &  $241.5\pm  14.5$& \cite{marek} \\ \hline
&                &&       \\
S + S 200 A$\cdot$GeV&$171.0$         &$160.04$& \cite{spieles}  \\ \hline
&           &&        \\
S + S 200 A$\cdot$GeV&$160.2\pm 3.5$   &  $158.0\pm 4.0$&\cite{sollfrank}  \\ \hline
&         &&     \\
S + S 200 A$\cdot$GeV&$165.0\pm    5.0$    &$175.0\pm 5.0$&\cite{stachel2} \\\hline\hline
AGS  &         &&         \\ \hline
     &         &&       \\
Si + Au 14.6 A$\cdot$GeV     & $127\pm 8$      &$485\pm 70$& \cite{tounsi}    \\ \hline
     &         &&       \\
Si + Au 14.6 A$\cdot$GeV     & $130\pm 10$      &$540\pm 20$& \cite{stachel1}    \\ \hline
     &            &&      \\ 
Si + Au 14.6 A$\cdot$GeV     & $118\pm 12$      &$522\pm 58$& \cite{becattini2}    \\ \hline
     &            &&      \\ 
 Si + Au 14.6 A$\cdot$GeV    & $110\pm 5$ &$540\pm 20$  & \cite{elliott} \\ \hline\hline
SIS  &         &&      \\  \hline
     &         &&       \\
Ni-Ni 1.9 A$\cdot$GeV & $70\pm 10$ &$720\pm 50$& \cite{keranen,prc} \\ \hline
     &         &      &      \\
Ni-Ni 0.8 A$\cdot$GeV & $48\pm 10$  &$820\pm 10$& \cite{prc}\\\hline
&             &           &      \\
Au-Au 1.0 A$\cdot$GeV &  $49\pm 3$   &$825\pm 8$ &\cite{prc}   \\ \hline
&             &                 &\\
 Ni-Ni 1.0 A$\cdot$GeV & $51\pm 5$   &$822\pm 10$&\cite{prc}   \\ \hline
&             &                 &\\
 Ni-Ni 1.8 A$\cdot$GeV & $54\pm 3$   &$806\pm 8$ &\cite{prc}   \\ \hline
\end{tabular}
\caption{Chemical freeze-out temperature $T_{ch}$ and baryon
chemical potential $\mu_B^{ch}$ in various collisions.
}
\end{center}
\end{table}

%
\newpage
%
\begin{table}
\begin{center}
\begin{tabular}{|l|c|c|c|} 
Energy & $T_{f}$ [MeV] &  $\mu_B^{f}$ [MeV] & Reference\\
       &           &                &         \\ \hline \hline
SPS    &           &                &         \\ \hline
       &           &                &         \\
Pb + Pb 158 A$\cdot GeV$ &$120\pm 12$ &$300\pm 35$  & \cite{NA49}$^\dagger$ \\ \hline
       &           &                &         \\
Pb + Pb 158 A$\cdot GeV$ &$95.8\pm 3.5$ &  $400\pm 24$& \cite{nix2}$^\dagger$ \\ \hline
       &           &                &         \\
Pb + Pb 158 A$\cdot GeV$&$100$ &  $400$& \cite{tomasik} \\
\hline\hline
&                &&       \\
&         &&     \\
AGS  &         &&         \\ \hline
     &         &&       \\
Si + Au 14.6 A$\cdot$GeV     & $93.4\pm 4.4$      &$554\pm 35$& \cite{nix}    \\ \hline
     &            &&      \\ 
 Si + Au 14.6 A$\cdot$GeV    & $93$ &$540$  & \cite{dobler} \\ \hline
 
\end{tabular}
\caption{Thermal freeze-out temperature $T_f$ and baryon
chemical potential $\mu_B^f$ in various collisions.
$^\dagger $The value of $\mu_B^f$ was estimated  using isentropic 
expansion starting from the chemical freeze-out points.}
\end{center}
\end{table}

\newpage
\begin{table}
\begin{center}
\begin{tabular}{|c|c|cl|}
\hline
       & Thermal&             &      \\
       & Model  & exp. data   & Ref. \\
       & 40 GeV & 158 GeV     &  \\
       &        &             &      \\
\hline
\(\rm (p-\overline{p})/h^-\) & 0.382 & 0.228$\pm$ 0.029  & \cite{NA49}\\
\(\rm \overline{p}/p\)       & 0.006 & 0.055$\pm$ 0.010  & \cite{NA44}\\
\(\rm \Lambda/K^0_s\)        & 0.66  & 0.65$\pm$ 0.11   & \cite{NA49}\\
\(\rm K^+/K^-\)              & 2.34  & 1.85$\pm$ 0.09    & \cite{NA44}\\
\(\rm \overline{\Lambda}/\Lambda\) & 0.014 & 0.128$\pm$ 0.012 & \cite{WA97}\\
\(\rm \Xi^-/\Lambda\)        & 0.12  & 0.127$\pm$ 0.011  & \cite{WA97}\\
\(\rm \overline{\Xi^-}/\overline{\Lambda}\) & 0.24 & 0.180$\pm$ 0.039& \cite{WA97}\\
\(\rm \Xi^+/\Xi^-\)          & 0.038  & 0.227$\pm$ 0.033  & \cite{WA97}\\
\(\rm \overline{\Omega^-}/\Omega^-\)    & 0.14 & 0.46$\pm$ 0.15   & \cite{WA97}\\
\end{tabular}
\caption{Model Predictions for \(T=140\) MeV, \(\mu_B=380\) MeV
corresponding to beam energy of 40 GeV compared to experimental
results obtained at 158 GeV.}
\label{table2}
\end{center}
\end{table}
\end{document}